# THE FLUCTUATION THEOREM FOR

# STOCHASTIC SYSTEMS

Debra J. Searles

Department of Chemistry

University of Queensland

Brisbane, Qld 4072 Australia

Denis J. Evans

Research School of Chemistry

Australian National University

Canberra, ACT 0200 Australia

**ABSTRACT**

The Fluctuation Theorem describes the probability ratio of observing trajectories that satisfy or violate the second law of thermodynamics. It has been proved in a number of different ways for thermostatted deterministic nonequilibrium systems. In the present paper we show that the Fluctuation Theorem is also valid for a class of stochastic nonequilibrium systems. The Theorem is therefore not reliant on the reversibility or the determinism of the underlying dynamics. Numerical tests verify the theoretical result.

## 1. INTRODUCTION

The Fluctuation Theorem (FT) states that the ratio of the probability of observing nonequilibrium trajectory segments of duration $\tau$ with a time averaged rate of entropy production per unit volume $\overline{\sigma}_\tau$, to the probability of observing segments with an average entropy production rate per unit volume $-\overline{\sigma}_\tau$ is,

$$\ln\left(\frac{p(\overline{\sigma}_\tau)}{p(-\overline{\sigma}_\tau)}\right) = \frac{1}{k_B}\overline{\sigma}_\tau V\tau \qquad (1)$$

The FT is interesting in that it gives an analytic expression for the probability that, for a finite system and for a finite time, the Second Law of Thermodynamics will be violated. This expression has been tested numerically and predicts the expected long time and large system behaviour, that is: Second Law violating trajectories *will not be observed* in the thermodynamic and/or long time limits.

There are three approaches that have been used to derive this expression for deterministic systems. This relationship was derived in 1993 for nonequilibrium steady states by Evans, Cohen and Morriss (ECM2) [1] from a natural invariant measure [2] which was proposed heuristically for steady state trajectories. In 1995 Gallavotti and Cohen [3, 4] gave a proof of the Steady State FT, demonstrating that this FT can be derived from the Sinai-Ruelle-Bowen measure (SRB) [5], if one employs the so-called Chaotic Hypothesis [3,4]. The Steady State FT (1) is only true asymptotically, $\tau \rightarrow \infty$. In ECM2 it was shown [1] that the Steady State Fluctuation Theorem (1), was consistent with computer simulation results for an atomic fluid undergoing, reversible thermostatted shear flow far from equilibrium.

A Transient FT, had already been developed in 1994 [6,7,8]. It considers transient trajectories which are generated from initial phases sampled from an equilibrium microcanonical ensemble and which evolve in time towards the steady state which is assumed to be unique. Unlike Steady State FT's this Transient Fluctuation Theorem is exact for **all** trajectory durations, $\tau$. If long time steady state averages are independent of the initial phase vector used to generate a steady state trajectory segment (for a given volume, number of particles and energy), then one might expect that, in the limit $\tau \to \infty$, averages over transient trajectory segments would approach those taken over nonequilibrium steady state segments. Further, one would expect the asymptotic convergence of the Transient and Steady State Fluctuation Theorems. However, there has been some recent discussion of this point and not all parties agree about this asymptotic convergence[9].

A third approach that has been used to derive the FT is valid only in the linear response regime close to equilibrium and is also only valid asymptotically ($\tau \to \infty$). In a footnote to ECM2 [1], the authors noted that the FT can be proved using the Green-Kubo relations for the linear response together with an application of the Central Limit Theorem to the distribution of $\{ \bar{\sigma}_\tau \}$, in the $\tau \to \infty$ limit [7,10]. This third approach, although limited to the linear response regime, is quite general with respect to the nature of the thermostatting. In an obvious limit this approach applies to *unthermostatted* systems. This is because Green-Kubo relations are robust with respect to thermostatting (see p 116 *et. seq.* of [11]).

Most theoretical and numerical studies of the FT have concentrated on reversible, deterministic dynamics although recently theoretical studies on stochastic systems have been carried out [12,13,14]. Kurchan [12] has shown that the the FT is valid for Langevin dynamics, and Lebowitz and Spohn [13] showed that it could be extended to apply to steady state Markov processes. Maes recently demonstrated [14] a FT can be obtained if the steady state is regarded as a Gibbs state.

In the present paper, the Transient FT, is generalised so that it applies to stochastic systems. Furthermore, it is demonstrated that by considering the transient response of a system that is initially in a state with a known distribution function, rather than directly treating a steady state system, a formula that is valid at all times is obtained. This approach is different from the steady state approach of Lebowitz and Spohn [13] in that an exact, finite time Transient FT is derived. As in the deterministic case, if the steady state is unique, we expect that the Transient FT will asymptotically converge to the Steady State stochastic FT. Also, we provide the first numerical tests of a stochastic FT. Given that the FT is valid for stochastic systems, reversibility and determinism are clearly not prerequisites for the FT.

## 2. DERIVATION OF THE FT USING THE LIOUVILLE MEASURE FOR A CLASS OF STOCHASTIC SYSTEMS

Consider the equations of motion for a stochastic system given by:

$$\dot{\mathbf{\Gamma}} = \mathbf{G}(\mathbf{\Gamma}) + \mathbf{\xi}(t) \tag{2}$$

where $\mathbf{\xi}(t)$ is a random variable. The first term on the right hand side $\mathbf{G}(\mathbf{\Gamma})$, is deterministic and assumed to be reversible. As an example, consider the transient response of a system, initially at equilibrium, to an applied field with a random term, $\mathbf{\xi}_i$, contributing to the equations of motion for the momenta. The system is thermostatted to ensure a steady state can be reached and the equations of motion are:

$$\dot{\mathbf{q}}_i = \frac{\mathbf{p}_i}{m} + \mathbf{C}_i(\mathbf{\Gamma})F_e$$
$$\dot{\mathbf{p}}_i = \mathbf{F}_i(\mathbf{q}) + \mathbf{D}_i(\mathbf{\Gamma})F_e + \mathbf{\xi}_i(t) - \alpha\mathbf{p}_i \tag{3}$$

where $\mathbf{q}_i$ and $\mathbf{p}_i$ are the coordinates and momenta of the ith particle, respectively, $\mathbf{F}_i$ is the interparticle force on that particle, $F_e$ is an external field applied to the system, $\mathbf{C}_i$ and $\mathbf{D}_i$ describe the coupling of the system to the field, and $\alpha$ is a Gaussian thermostat multiplier [11] that fixes the internal energy:

$$\alpha = \frac{\sum_{i=1}^{N} F_e\mathbf{D}_i(\mathbf{\Gamma}) \cdot \mathbf{p}_i / m + \mathbf{\xi}_i \cdot \mathbf{p}_i / m - F_e\mathbf{C}_i(\mathbf{\Gamma}) \cdot \mathbf{F}_i(\mathbf{q})}{\sum_{i=1}^{N} \mathbf{p}_i \cdot \mathbf{p}_i / m}. \tag{4}$$

To ensure that the system remains on a constant energy, zero total momentum, hypersurface, the thermostat multiplier contains the random term and the restriction, $\sum_{i=1}^{N} \mathbf{\xi}_i = \mathbf{0}$, is imposed. The phase space of the nonequilibrium system is therefore is a subset of that of the initial equilibrium ensemble. In equation (3) the stochastic term can be regarded either as a random force that is added to the equation for the rate of change of momentum, or it can be regarded as contributing a random term to the thermostat. The difference between these two interpretations is purely semantic.

If the adiabatic incompressibility of phase space (AI$\mathbf{\Gamma}$) condition is satisfied [11], then the Liouville equation for this system reads:

$$\frac{df(\mathbf{\Gamma},t)}{dt} = -f(\mathbf{\Gamma},t)\frac{\partial}{\partial\mathbf{\Gamma}}\bullet\dot{\mathbf{\Gamma}} = -\Lambda(\mathbf{\Gamma})f(\mathbf{\Gamma},t) \qquad (5)$$

where $\Lambda(\mathbf{\Gamma})$ is the phase-space compression factor. For the system described by equations (3) and (4), $\Lambda(\mathbf{\Gamma}) = -dN\alpha(\mathbf{\Gamma}) + O(1)$ where d is the number of Cartesian coordinates considered. The solution of equation (5) can be written [7]

$$f(\mathbf{\Gamma}(t),t) = \exp[-\int_0^t \Lambda(s)ds]f(\mathbf{\Gamma},0). \qquad (6)$$

The FT considers the probabilities of observing trajectories with entropy production rates which are equal in magnitude but opposite in sign. In the proof of this theorem using the Liouville measure [6,7,8,10] it was necessary, for every possible trajectory, to identify a conjugate trajectory which had this property: that is all trajectories were sorted into conjugate pairs. In a reversible, deterministic system this identification was straightforward and was accomplished by carrying out time reversal mappings [6,7,8,10]. It is now shown how this procedure can be modified for stochastic systems.

Consider a trajectory segment, $\mathbf{\Gamma}_+(s)$; $0<s<t$ and its time-reversed trajectory, $\mathbf{\Gamma}_-(s)$; $0<s<t$ which we call an antisegment. The sign in the subscript reflects the sign of the integral of the thermostat multiplier (or entropy production) along the trajectory segment. For a reversible system, these trajectories are simply related by a time reversal mapping: each conjugate trajectory $\mathbf{\Gamma}_-$ is generated from the original trajectory $\mathbf{\Gamma}_+$ by carrying out a time-reversal mapping of the phase at the midpoint of the trajectory and integrating the equations of motion backward and forward in time [6,7,8,10]. Without loss of generality if the field is assumed to be even with respect to the time-reversal mapping, then the flux, entropy production rate and the thermostat multiplier will be odd. We use the notation that the averages of a phase variable, A, along a forward trajectory and its conjugate, time reversed, trajectory are given by:

$$\overline{A}_+(t) \equiv \frac{1}{t} \int_0^t ds\ A(\mathbf{\Gamma}_+(s)) \qquad\qquad (7)$$

and

$$\overline{A}_-(t) \equiv \frac{1}{t} \int_0^t ds\ A(\mathbf{\Gamma}_-(s)) \qquad\qquad (8)$$

respectively. Depending on the parity of the phase function $A(\mathbf{\Gamma})$ under time reversal symmetry, there may be a simple relation between $\overline{A}_+(t)$ and $\overline{A}_-(t)$.

In a stochastic system the conjugate trajectory can no longer be generated by simply carrying out a time reversal mapping and solving the equations of motion. After the time reversal mapping at the midpoint of the original trajectory, integration of the equations of motion forward and backward in time will with overwhelming probability, result in the observation of a different set of random numbers than were observed for the original trajectory and the trajectories will not be conjugate. Clearly a mapping of the sequence of random numbers observed for the forward trajectory must be carried out for the conjugate trajectory. The necessary mapping of the random numbers will depend on the function $\xi_i(R)$ where R is a random number. Figure 1 give a diagrammatic representation of the way in which conjugate trajectories are generated for stochastic systems.

If the sequence of random numbers: $R_1$, $R_2$, $R_3$, $R_4$ is observed for the original trajectory, then this sequence must be appropriately mapped in the conjugate trajectory. For example if the random term contributes only to the equation of motion for $\dot{p}_{xi}$, which is even under a time-reversal mapping, then the sequence $R_4$, $R_3$, $R_2$, $R_1$ must be observed for the conjugate trajectory to be generated. Similarly, if the random term contributes only to $\dot{q}_{xi}$, which is odd under time reversal, then the sequence -$R_4$, -$R_3$, -$R_2$, -$R_1$ must be observed. Provided the mapped sequence is allowed by the random number generator, the antisegment will be a solution of the equations of motion. It should be noted that in the first case no restrictions on the random number generator are required however in the second case, a symmetry restriction on the range of numbers is necessary for this proof to be valid.

It can be observed from Figure 1 that $M^{(T)}\boldsymbol{\Gamma}_{(3)} = \boldsymbol{\Gamma}_{(4)}$; $M^{(T)}\boldsymbol{\Gamma}_{(1)} = \boldsymbol{\Gamma}_{(6)}$ and $M^{(T)}\boldsymbol{\Gamma}_{(2)} = \boldsymbol{\Gamma}_{(5)}$ where is $M^{(T)}$ used to to represent a time-reversal mapping. At all points along the trajectory the fluxes of conjugate trajectories are related by $J(t;\boldsymbol{\Gamma}_{+}, 0 < t < \tau) = -J(t;\boldsymbol{\Gamma}_{-}, 0 < t < \tau)$ and therefore fluxes averaged over the duration of the segment are related by: $\bar{J}_{+} = -\bar{J}_{-}$. It is straightforward then to see that in order to divide the trajectories into conjugate pairs, the equations of motion do not have to be reversible (that is $M^{(T)} \cdot e^{iL(\Gamma)t} \cdot M^{(T)} \cdot e^{iL(\Gamma)t} \cdot \boldsymbol{\Gamma}(0) = \boldsymbol{\Gamma}(0)$ where $M^{(T)} \cdot M^{(T)} = 1$), but it is necessary that *the antisegment is a solution of the equations of motion*. This condition is equivalent to that required in the derivation by Lebowitz and Spohn [13] in which case it is assumed that if the rate constant for a forward step is non-zero, then the rate of the reverse process must also be non-zero. That is, in both derivations it is required that the reverse process is able to be observed.

Now it is shown how the probability of observing the conjugate trajectories can be determined. For the system considered, the initial phases are distributed microcanonically, so the probability of observing an initial phase inside a small phase volume, $\delta V(\boldsymbol{\Gamma}(0))$ about $\boldsymbol{\Gamma}(0)$ is proportional to $\delta V(\boldsymbol{\Gamma}(0)$. It is assumed that our universe is causal: the probability of observing a trajectory segment is proportional to the probability of observing the *initial* phase that generates the segment. Using the fact that for sufficiently small volumes, $\delta V(\boldsymbol{\Gamma}(t) \sim 1/f(\boldsymbol{\Gamma}(t)$ and that the Jacobian for the time reversal mapping is unity, the solution the Liouville equation given by equation (5) allows the expansion or contraction of a phase volume along a trajectory to be determined by,

$$V(\boldsymbol{\Gamma}(t), t) = \exp[\int_0^t \Lambda(\boldsymbol{\Gamma}(s))ds]V(\boldsymbol{\Gamma}, 0) \qquad (9)$$

This is illustrated in Figure 2.

The ratio of volumes the $\delta V_1$ and $\delta V_4$ gives the ratio of the probability of observing *initial phase*

*points*. The probability of observing a *trajectory* is equal to the product of the probability of observing the initial phase point and the probability of observing the sequence of random numbers:

$$\text{prob}(\Gamma(s); 0 < s < t) = \text{prob}(\delta V(0))\text{prob}(R_1 \text{K } R_n) \tag{10}$$

The probability of observing a trajectory segment with a particular time-averaged value of $\Lambda$ is then given by the sum over all trajectories with that value, and the probability ratio is given by:

$$\frac{p(\overline{\Lambda}_+(\tau))}{p(\overline{\Lambda}_-(\tau))} = \frac{\displaystyle\sum_{i\left|\int\Lambda(\Gamma_i)=\overline{\Lambda}_+(\tau)\right.}\delta V(\Gamma_{i+}(0))p(R_1...R_n)_i}{\displaystyle\sum_{i\left|\int\Lambda(\Gamma_i)=\overline{\Lambda}_-(\tau)\right.}\delta V(\Gamma_{i+}(0))p(R_1...R_n)_i}$$

$$= \frac{\displaystyle\sum_{i\left|\int\Lambda(\Gamma_i)=\overline{\Lambda}_+(\tau)\right.}\delta V(\Gamma_{i+}(0))p(R_1...R_n)_i}{\displaystyle\sum_{i\left|\int\Lambda(\Gamma_i)=\overline{\Lambda}_+(\tau)\right.}\delta V(\Gamma_{i-}(0))p(M^{(T)}(R_1...R_n)_i)}$$

$$\tag{11}$$

$$= \frac{\displaystyle\sum_{i\left|\int\Lambda(\Gamma_i)=\overline{\Lambda}_+(\tau)\right.}\delta V(\Gamma_{i+}(0))p(R_1...R_n)_i}{\displaystyle\sum_{i\left|\int\Lambda(\Gamma_i)=\overline{\Lambda}_+(\tau)\right.}\exp[\overline{\Lambda}_{+i}(\tau)\tau]\delta V(\Gamma_{i+}(0))p(M^{(T)}(R_1...R_n)_i)}$$

$$= \exp[-\overline{\Lambda}_{+i}(\tau)\tau]$$

where the notation: $i\left|\int\Lambda(\Gamma_i) = \overline{\Lambda}_+(\tau)\right.$ is used to represent all trajectories, i, for which the time averaged value of the phase-space compression factor is equal to $\overline{\Lambda}_+$; and it is assumed that $p(R_1,...R_n) = p(M^{(T)}(R_1...R_n))$. The resulting fluctuation formula given by equation (11) for this stochastic system is identical to that for the deterministic, reversible systems. As in the deterministic case there may be many different pairs of conjugate trajectories which each have the same value for $\overline{\Lambda}_+(\tau)$.

The FT derived above is valid for *transient* trajectory segments of arbitrary length. Averages of phase variables over the transient trajectory segments approach the averages over steady state trajectory segments in the long time limit, therefore the stochastic FT will also apply to steady state systems.

# 3. NUMERICAL TESTS OF THE FLUCTUATION THEOREM APPLIED TO TRANSIENT STOCHASTIC SYSTEMS

Transient NEMD simulations of Couette flow using the SLLOD algorithm and the usual Lees-Edwards periodic boundary conditions, were carried out employing a *stochastic* force in the x-direction and the corresponding Gaussian isoenergetic thermostat. The equations of motion for this system are:

$$\dot{\mathbf{q}}_i = \mathbf{p}_i + \mathbf{i}\gamma y_i$$

$$\dot{\mathbf{p}}_i = \mathbf{F}_i - \mathbf{i}\gamma p_{yi} + \mathbf{i}\xi_i - \alpha\mathbf{p}_i \tag{12}$$

with the thermostat multiplier given by,

$$\alpha = \frac{\sum_{i=1}^{N}\xi_i p_{xi} - \gamma(p_{xi}p_{yi} + F_{xi}y_i)}{\sum_{i=1}^{N}\mathbf{p}_i \cdot \mathbf{p}_i} \tag{13}$$

The system consisted of N = 32 particles in 2 Cartesian dimensions and the particles interacted with the WCA short ranged, repulsive pair potential [15]. Lennard-Jones units are use throughout. The internal energy per particle was set at E/N = 1.56032 (*i.e.* T~1.0) and the particle density at n = N/V = 0.8. A strain rate of $\gamma = 0.5$ was applied.

Since for this system, $\Lambda(\mathbf{\Gamma}) = -2N\alpha(\mathbf{\Gamma}) + O(1) = \sigma(\mathbf{\Gamma})V / k_B$, the fluctuation theorem becomes,

$$\frac{p(\overline{\alpha}_+(\tau))}{p(\overline{\alpha}_-(\tau))} = \exp[2N\overline{\alpha}_+(\tau)\tau] \tag{14}$$

where O(1) are omitted since they are negligible in the thermodynamic limit, and to reduce the complexity of the expression. The system studied here is sufficiently small that these effects cannot be neglected and they are included in the data presented.

The stochastic term is of the simplest kind: the sequence of random numbers that must be observed in order to generate a conjugate trajectory are the same numbers that are observed for the forward trajectory, but they must occur in the reverse order. This means that no restrictions on the distribution of the random numbers are required to obtain equation (14). In the simulations, the stochastic term, $\xi_i$, was the product of a random number and a delta function at each time step. The random numbers were selected from a Gaussian distribution with zero mean, a standard deviation of 1.0 and were restricted within the range [-10.0,10.0].

Figure 3 shows the ensemble averaged response of the flux which indicates that a steady state is approached and the initial transient response has a Maxwell time of approximately $\tau_M = 0.07$.

In the transient response simulations, many initial equilibrium phases were generated and the response to an applied strain rate was monitored for various trajectory segment lengths. Histograms of $\overline{\alpha}_+(\tau) \equiv \frac{1}{\tau} \int_0^\tau ds\, \alpha(\mathbf{\Gamma}_+(s))$ were obtained with $\tau = 0.1$, 0.4 and 0.6. These histograms are shown in Figure 4. The FT predicts that a plot of $\left(\ln\!\big(p(\overline{\alpha}_+(\tau))/p(-\overline{\alpha}_+(\tau))\big)\right)/(2N\tau)$ versus $\overline{\alpha}_+(\tau)$ should give a straight line of unit slope. For each of the trajectory segment lengths considered in Figure 3 the FT was tested with O(1/N) corrections included. The normalised probability ratios are shown in Figure 5. In each case a slope of unity is obtained and the FT is verified.

These results show that the FT is valid for finite averaging times of this transient, stochastic system.

# 4. NUMERICAL TESTS OF THE FLUCTUATION THEOREM APPLIED TO STEADY STATE STOCHASTIC SYSTEMS

The FT was also examined for steady state systems evolving with the stochastic equations of motion considered in Section 3. Histograms of $\overline{\alpha}_+(\tau) \equiv \frac{1}{\tau}\int_0^\tau ds\, \alpha(\Gamma_+(s))$ for steady state trajectory segments of length $\tau = 0.05, 0.1, 0.2, 0.3$ and $0.4$ were calculated. The results for $\tau = 0.05, 0.2$ and $0.4$ are shown in Figure 6.

For steady state trajectories, the FT predicts that a plot of $\left(\ln\left(p(\overline{\alpha}_+(\tau))/p(-\overline{\alpha}_+(\tau))\right)\right)/(2N\tau)$ vs $\overline{\alpha}_+(\tau)$ gives a straight line of unit slope *in the limit* $\tau\rightarrow\infty$. Figure 7 shows the results and figure 8 plots the slope as a function of $\tau$, indicating that it approaches unity in the long time limit.

## 5. CONCLUSIONS

The present work shows that the Fluctuation Theorem is quite general and applies to both deterministic and stochastic nonequilibrium systems. As was found to be the case for deterministic systems, the Fluctuation Theorem applies:

•        at all times to finite (transient) trajectory segments which are initially sampled from the equilibrium microcanonical ensemble and then move isoenergetically towards a steady state, and,

•        asymptotically to long time steady state trajectory segments. In all cases, transient or steady state, stochastic or deterministic, the Fluctuation Theorem applies in both the linear and the nonlinear response regimes.

As a final comment we note that although the theory and simulations presented here apply to systems in which every particle is ergostatted (so-called homogeneous thermostatting), the theory presented here applies equally well to systems were only a subset of the particles are thermostatted [16]. The theory also applies to systems composed of mixtures of particles with different interparticle interactions. We can therefore obviously model boundary thermostatted systems where a fluid obeying Newtonian mechanics (ie no thermostat) flows inside thermostatted solid walls, using the theory presented here. To treat such a system you consider a mixture of two types of particles where at the temperature and density studied one set of particles, the wall particles, are in the solid phase and are thermostatted and the other set of particles are liquid and are not thermostatted. In such cases the only difference to the theory above is that in equations such as (6), above the N, refers to the number of thermostatted particles and not to the total number of particles.

### Acknowledgements

We would like to thank the Australian Research Council for the support of this project. Helpful discussions and comments from Professor E.G.D. Cohen are also gratefully acknowledged.

**Figure Captions.**

Figure 1. A schematic diagram showing the construction of conjugate trajectories for stochastic systems.

Figure 2. A schematic diagram representing the change in phase volume with time along a trajectory and its conjugate.

Figure 3. The ensemble averaged response of the flux for as system of 32 particles in 2 Cartesian dimensions to which a strain rate is applied at time zero and for which the SLLOD algorithm is used to model the shear flow. The internal energy per particle was set at E/N = 1.56032 (*i.e.* T~1.0) and the particle density at n = 0.8. A strain rate of $\gamma = 0.5$ was applied.

Figure 4. Histograms of $\overline{\alpha}_+(\tau) \equiv \frac{1}{\tau}\int_0^\tau ds\ \alpha(\Gamma_+(s))$ for a system undergoing transient response to an applied strain rate of $\gamma = 0.5$. The internal energy per particle was set at E/N = 1.56032 (*i.e.* T~1.0) and the particle density at n = 0.8. Trajectory segments of a) $\tau = 0.1$, b) $\tau = 0.4$ and c) $\tau = 0.6$ were used.

Figure 5. Plots of $\left(\ln\left(p(\overline{\alpha}_+(\tau))/p(-\overline{\alpha}_+(\tau))\right)\right)/(2N\tau)$ vs $\overline{\alpha}_+(\tau)$ for the system considered in Figure 4 with a) $\tau = 0.1$, b) $\tau = 0.4$ and c) $\tau = 0.6$. Order (1/N) corrections are included. The straight line is of unit slope and it is the result predicted from the FT. The slopes obtained from weighted least squares fits are a) 0.98±0.01; b) 1.00±0.02; c) 1.02±0.04.

Figure 6. Histograms of $\overline{\alpha}_+(\tau) \equiv \frac{1}{\tau}\int_0^\tau ds\ \alpha(\Gamma_+(s))$ for a system undergoing steady state shear flow with an applied strain rate of $\gamma = 0.5$. The internal energy per particle was set at E/N = 1.56032 (*i.e.* T~1.0) and the particle density at n = 0.8. Trajectory segments of a) $\tau = 0.05$, b) $\tau = 0.2$ and c) $\tau = 0.4$ were used.

Figure 7. The slope of plots of $\left(\ln\left(p(\overline{\alpha}_+(\tau))/p(-\overline{\alpha}_+(\tau)))\right)/(2N\tau)$ vs $\overline{\alpha}_+(\tau)$ for various trajectory segment lengths. The result is consistent with a convergence to a value of unity in the long time limit which is the result predicted from the FT.

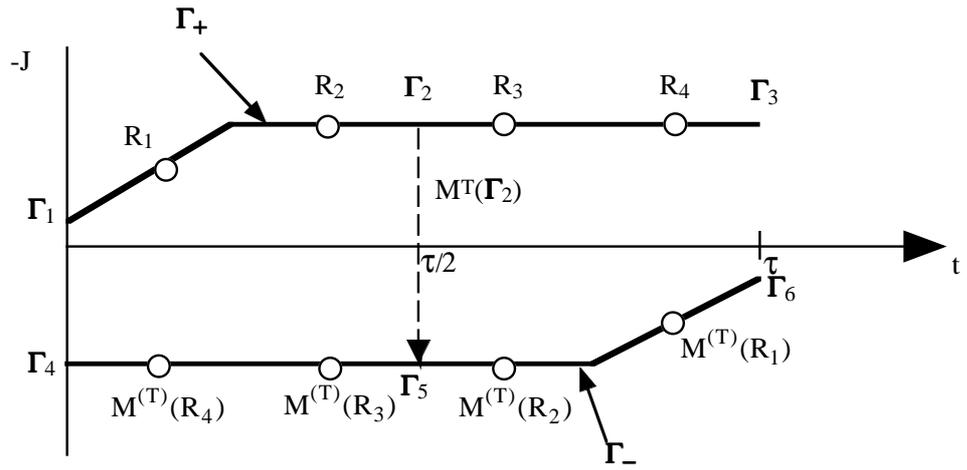

Figure 1

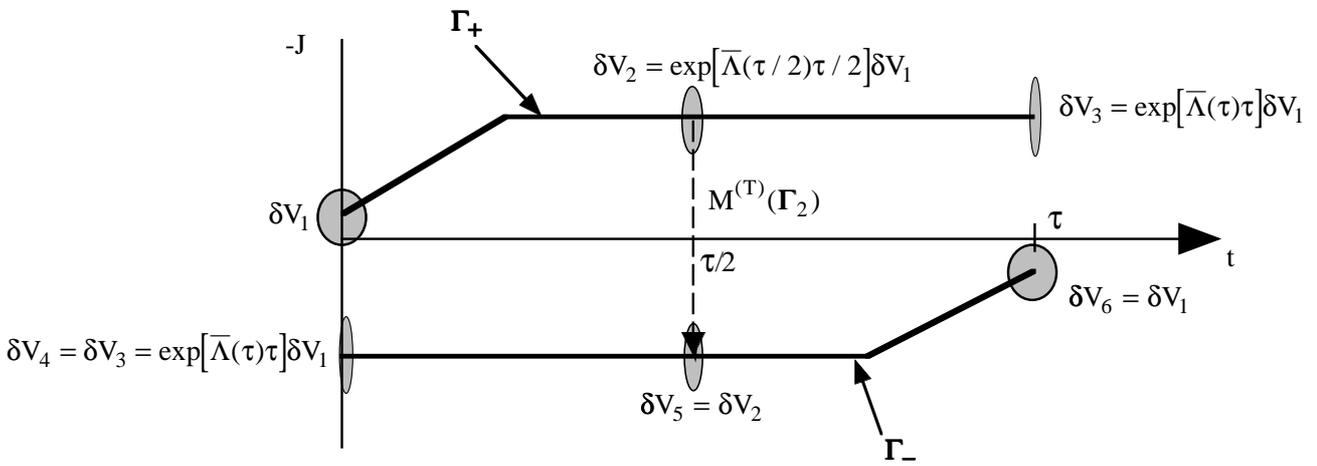

Figure 2

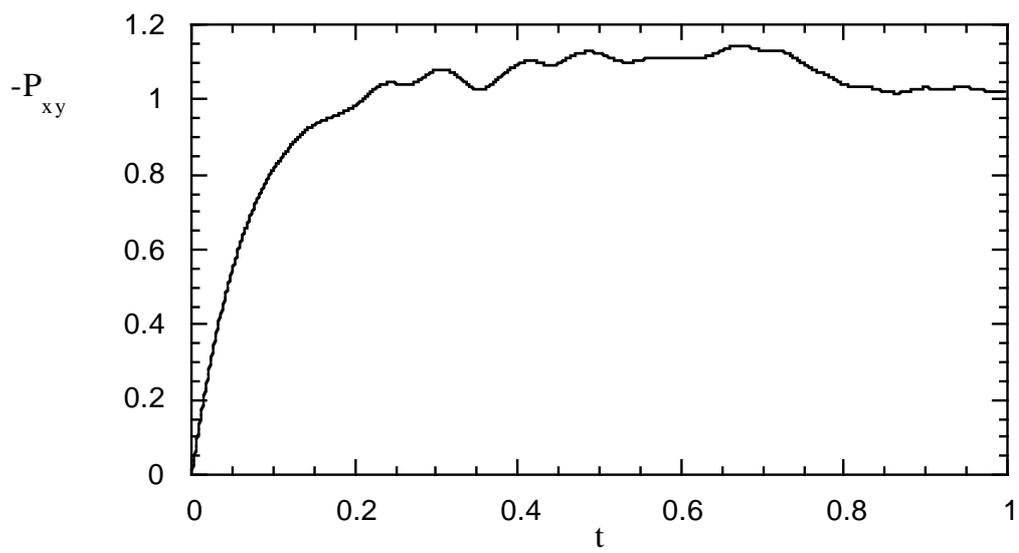

Figure 3

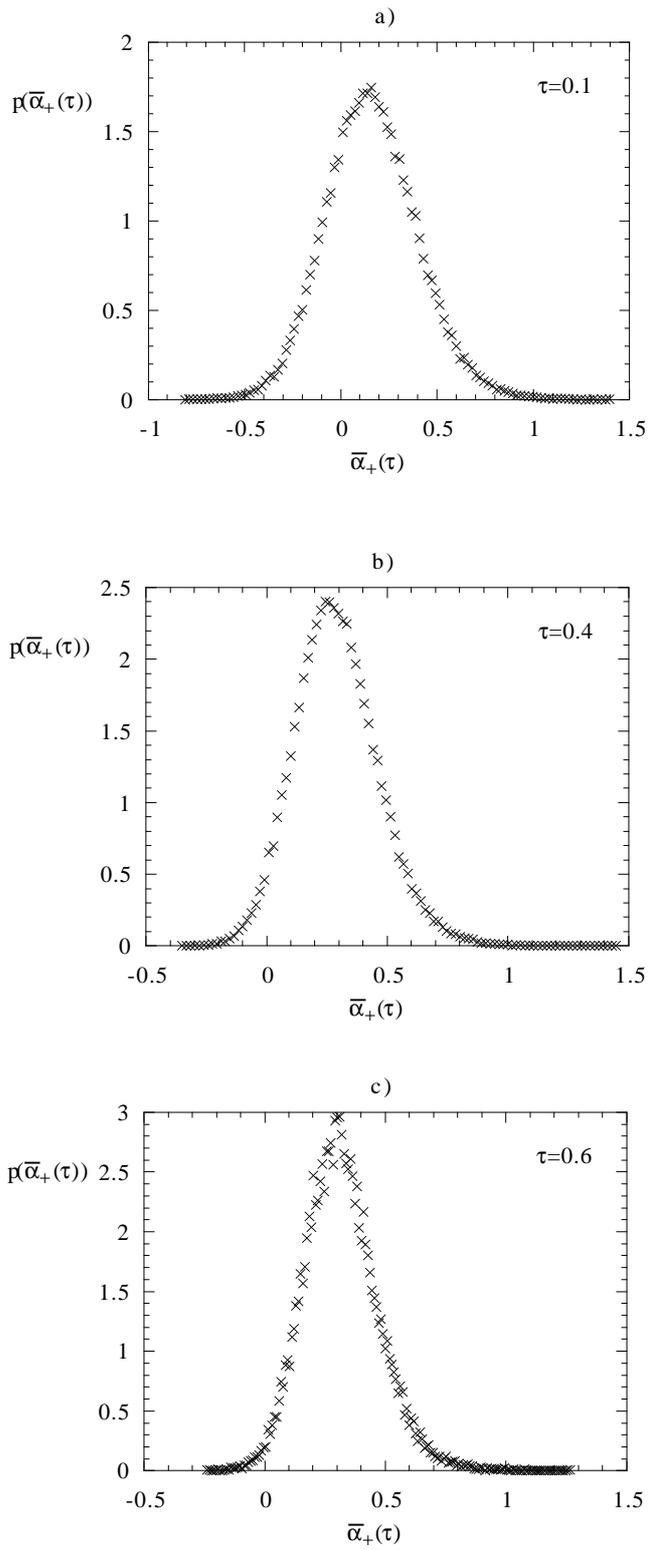

Figure 4

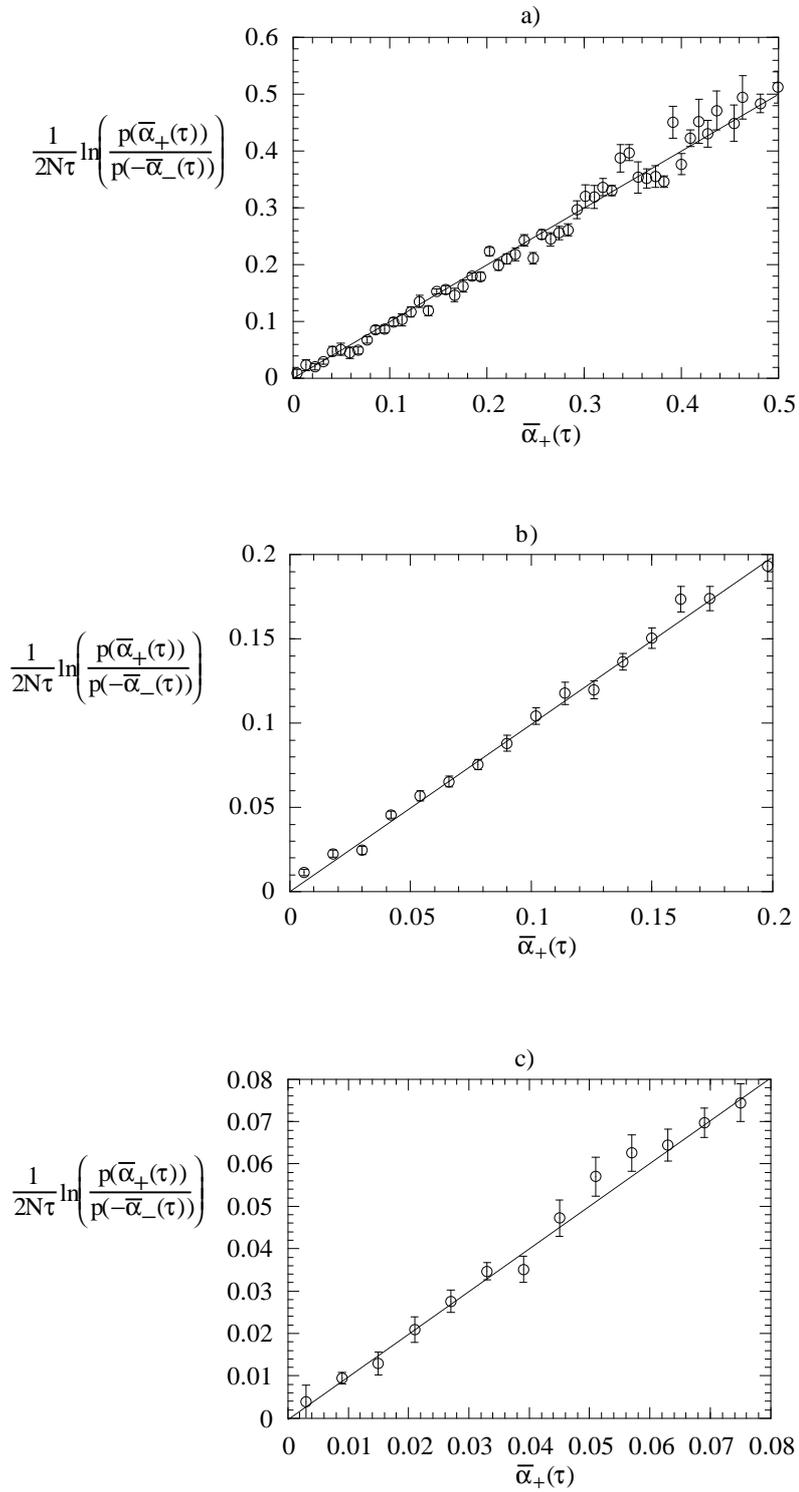

Figure 5

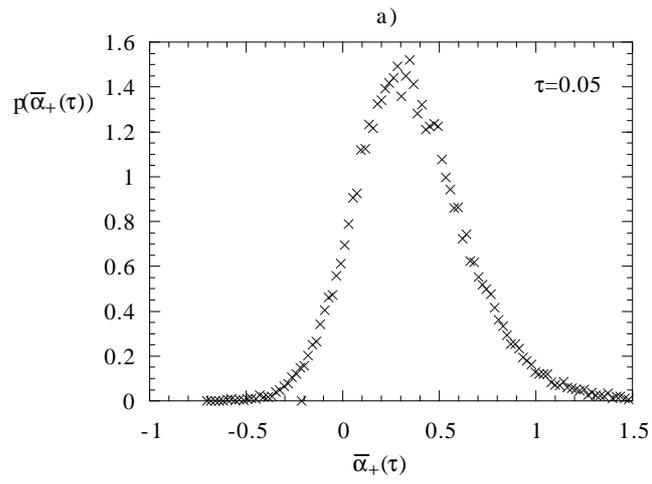

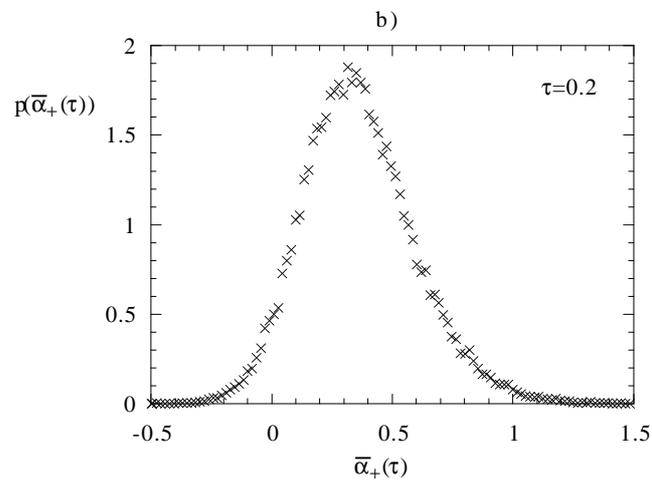

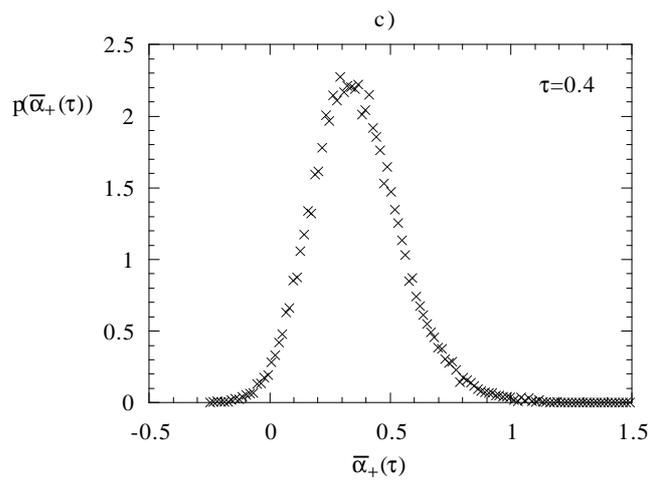

Figure 6

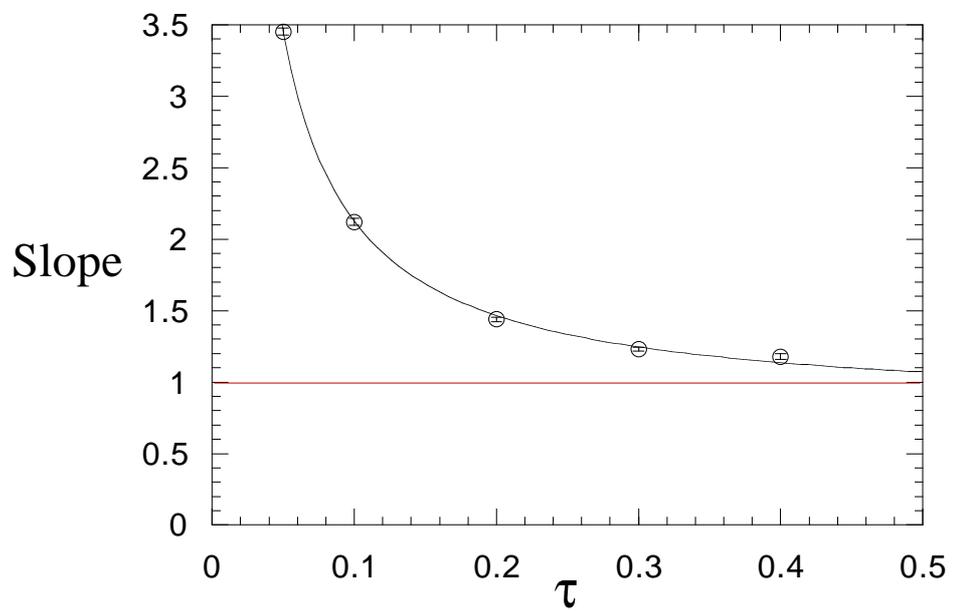

Figure 7